\documentclass[12pt,english]{article}
\usepackage{epsfig, amssymb, graphicx, amsmath}

\usepackage{wrapfig}	
\usepackage{float}		
\usepackage{subfig}	
\usepackage{footnote}
\usepackage{mathtools}
\usepackage{tabularx}
\usepackage{booktabs}

\usepackage{bm}
\usepackage[authoryear]{natbib} 
\bibpunct{(}{)}{,}{a}{}{,}

\newtheorem{remark}{Remark}

\newtheorem{proposition}{Proposition}

\newtheorem{definition}{Definition}
\newtheorem{lemma}{Lemma}
\newtheorem{corollary}{Corollary}

\newcommand{\E}{\mathbb{E}}
\newcommand{\V}{\mathbb{V}}

\newcommand{\be}{\begin{equation}}
\newcommand{\en}{\end{equation}}
\newcommand{\ben}{\begin{equation*}}
\newcommand{\enn}{\end{equation*}}
\newcommand{\bea}{\begin{eqnarray}}
\newcommand{\ena}{\end{eqnarray}}

\DeclareMathOperator{\Erfi}{Erfi}
\DeclareMathOperator{\Erf}{Erf}
\DeclareMathOperator{\Erfid}{Erfid}

\begin{document}

\newlength\tindent
\setlength{\tindent}{\parindent}
\setlength{\parindent}{0pt}
\renewcommand{\indent}{\hspace*{\tindent}}

\begin{savenotes}
\title{ Stop-loss and Leverage \\ in optimal Statistical Arbitrage with \\  an application to Energy market}
\author{
Roberto Baviera$^\dagger$ \&  
Tommaso Santagostino Baldi$^\ddagger$
}

\maketitle

\vspace{0.1cm}

\begin{tabular}{ll}
$(\dagger)$ &  Politecnico di Milano, Department of Mathematics, 32 p.zza L. da Vinci, I-20133 Milano \\
$(\ddagger)$ & J.P. Morgan Chase Bank, 25 Bank St., Canary Wharf, London E14 5JP\footnote{The views expressed here are those of the author and not necessarily those of the bank.}
\end{tabular}
\end{savenotes}

\vspace{0.1cm}

\begin{abstract}
In this paper we develop a statistical arbitrage trading strategy with two key elements in hi-frequency trading: stop-loss and leverage. 
We consider, as in \citet{bertram1}, a mean-reverting process for the security price with proportional transaction costs;
we show how to introduce stop-loss and leverage in an optimal trading strategy.

\noindent
We focus on repeated strategies using a self-financing portfolio.
For every given stop-loss level we derive analytically the optimal investment strategy consisting of optimal leverage and market entry/exit levels.

\noindent
First we show that the optimal strategy {\it \'a la} Bertram depends on the probabilities to reach entry/exit levels, on expected First-Passage-Times and 
on expected First-Exit-Times from an interval.
Then, when the underlying log-price follows an Ornstein-Uhlenbeck process,  we deduce analytical expressions for expected First-Exit-Times and
we derive the long-run return of the strategy as an elementary function of the stop-loss.

\noindent
Following industry practice of pairs trading we consider an example of pair in the energy futures' market, 
reporting in detail the analysis for a spread on Heating-Oil and Gas-Oil futures in one year sample of half-an-hour market prices.
\end{abstract}

\vspace*{0.11truein}
\vspace*{0.11truein}

\vspace{0.3cm}

\begin{flushleft}
{\bf Address for correspondence:}\\
Roberto Baviera\\
Department of Mathematics \\
Politecnico di Milano\\
32 p.zza Leonardo da Vinci \\ 
I-20133 Milano, Italy \\
Tel. +39-02-2399 4630\\
Fax. +39-02-2399 4621\\
roberto.baviera@polimi.it
\end{flushleft}

\newpage

\begin{center}
{\bf\Large Stop-loss and Leverage in optimal Statistical Arbitrage \\
with an application to Energy market}
\end{center}

\section{Introduction}
Statistical arbitrage trading 
exploits some statistical regularities in the dynamics of security prices,
generally implementing repeated strategies; in this way it
is able to obtain, with a high probability, a return larger than the risk-free return. 
These strategies are widespread
among hedge funds. 

A quantity of interest for an investor is strategy return per  unit time up to time $t$ (hereinafter return)
\[
 \mu_t \coloneqq \frac{1}{t} \ln \frac{W_t}{W_0} \, ,
\]
where  $W_t$ is investor's wealth invested in the strategy at time $t$.

We consider in particular the case in which the investor has the possibility 
to repeat the strategy an indefinite number of times: a common assumption within statistical arbitrage techniques. 
In the case of a repeated game with infinite horizon, 
an objective function that makes sense maximizing is the long-run return that is an almost sure quantity  \citep[see e.g.][]{kelly}.

This problem has been first tackled by \citet{taksar} who considered a portfolio with two assets with 
proportional costs, with the risky security that follows a Geometric Brownian Motion.
This study and the literature that has followed \citep[see e.g.][for a review]{guasoni} have clarified that two are the relevant questions for the investor:
i) when he needs to rebalance his portfolio and ii) what fraction of wealth should be invested in the risky security.

In real situations, this chosen dynamics for the  risky security appears to be oversimplified, 
and mean reversion is a relevant stylized fact that cannot be neglected  
especially in pairs trading \citep[see e.g.][]{elliott, gatev, vidyamurthy}. 
Recently \citet{bertram1} has considered the case where the log-price of  risky asset follows 
an Ornstein-Uhlenbeck process (hereinafter OU) with transaction costs proportional to security prices. 
He considers an ansatz for the selection of the optimal long-run strategy
identifying two trading bands where 
the investor has to buy and to sell, respectively, one unit of the security.
The optimally chosen strategy within this ansatz can generate, as shown by \citet{bucca}, significant returns
in commodity pairs trading in energy market; they show that these results can be achieved  even considering daily end-of-the-day prices. 

Unfortunately, when considering real statistical arbitrage strategies, 
even this approach is too simple.
A statistical arbitrage strategy always requires a stop-loss, that is a level where 
the investor closes out a losing position (negative scenario) in order to limit maximum loss. 
The presence of a stop-loss is fundamental when applying these strategies in practice: 
indeed one can never be sure that, for a given security, the statistical description observed in the past, persists in the future. 

Moreover hedge funds have the possibility to leverage their position, e.g. 
borrowing either from his primary broker or from other lenders. 
In this paper we model this problem providing closed form solutions. 

\bigskip

The main contributions of this paper to the existing literature on statistical arbitrage are twofold.

A first contribution is to solve analytically, for any given stop-loss level, the problem 
with one risky asset {\it \'a la} Bertram with the possibility to borrow at a fixed rate $r$:
we identify the optimal bands and the optimal leverage. 
This approach clarifies also that the long-run return  depends only on the probabilities to reach the trading bands  
and on the expected time interval between two trades in the strategy, time interval that is a simple function of a First-Passage-Time (hereinafter also FPT) and a  First-Exit-Time (FET).

Second, we deduce for an OU an analytical expression for expected FET from an interval via elementary functions. 
This allows a fast computation of the optimal strategy for any given stop-loss level, 
since analytical expressions of the probabilities to reach interval endpoints and expected FPTs
are already known in the literature \citep[see e.g.][]{Borodin, NRS}.


\bigskip

The paper is organized as follows. In Section {\bf 2} we describe the key elements in the proposed strategy and 
we deduce the long-run return   
as a function of stop-loss, trading bands and leverage with a continuous price process for the risky security; 
we also show how this quantity is linked to a Kullback-Leibler divergence \citep{kullback} associated 
to the optimal allocation problem.
In Section {\bf 3} we deduce, for an OU, an analytical expression for the expected First-Exit-Time from an interval
and specify the analytical expression for the long-run return in this case.
In Section {\bf 4} we apply the trading strategy to a pair in the energy market on half-an-hour time-series, presenting some numerical results.
We calibrate OU parameters via an In-Sample maximum likelihood discussing in detail the statistical significance of the estimated quantities,
we compute optimal trading bands for different levels of leverage and then 
we measure, in an Out-of-Sample period, the observed return. 
Section {\bf 5} concludes.
	
\section{Statistical Arbitrage 
in presence of Stop-Loss and Leverage}
\label{sec:TradingStrategy}

We consider 
an investor that can include a risk-free asset paying an interest rate $r$
and a risky security.
As in \citet{bertram1}, log-price dynamics for the risky security follows an OU, i.e. $p_t=e^{X_t}$ where
\be
dX_t= \kappa (\eta - X_t)\,dt + \sigma\,dB_t \,
\label{eq:OU complete}
\en
with $B_t$ a 1-dimensional Brownian motion while $\kappa, \sigma \in \Re^+$ and $\eta \in \Re$ are constant parameters:
$\sigma$ is the instantaneous standard deviation, $\kappa$ the rate of reversion and $\eta$ the mean-reverting value. 
As in \citet{bertram1} transaction costs are proportional to the price of the risky security 
with $c$ a positive constant modeling the total transaction cost associated with a trade.
This model set up describes in a simple way a situation well studied in pair trading, leading to strategies that are often 
named ``reversal statistical arbitrage". 

\bigskip

A first difference with \citet{bertram1} is that we consider a risk-free asset in the portfolio and the possibility for an optimal allocation, instead of allowing an investment in the risky asset only. 
A trading strategy is a sequence of individual trades performed on the continuous time stochastic process (\ref{eq:OU complete}).
In particular we focus on an
investor that can diversify his investment between these two assets in a self-financing strategy; 
as in Bertram, we consider a repeated strategy and maximize strategy's long-run return
\[
\mu \coloneqq \lim_{t \to \infty} \mu_t \, .
\]

We show that this approach allows to introduce leverage in an elementary way. 


\begin{definition}[Leverage]
Leverage is defined as the ratio at time $t$ between the amount invested in the risky asset and 
investor's total wealth $W_t$.
\end{definition}
 
In particular we consider leverage at times when trades take place.
The other difference w.r.t. Bertram's strategy is that we introduce a stop-loss.

\begin{definition}[Stop-loss]
Stop-loss is defined as the level $L$ for the log-price of the risky security 
such that, when it is reached,
the trading strategy is interrupted in any case incurring in a loss.
Before entering in a trade, the investor fixes a stop-loss level.
\end{definition}

This approach mimics the behaviour often observed in practice in statistical arbitrage strategies: 
the investor establishes {\it ex ante} a stop-loss level $L$ before starting the trading strategy.
Given dynamics (\ref{eq:OU complete}) with time independent parameters, a reasonable choice for
the stop-loss level $L$ is to consider a constant value 
during the whole strategy.

Once $L$ is fixed, the investor has to choose the optimal strategy given that constraint.
This technique reminds the one introduced in \citet{LeungLi}, where the stop-loss is introduced as a constraint for the strategy; 
they show how to determine the optimal bands for the expected earning of one single trade when security price is an OU.  
Unfortunately it is not straightforward to generalize their technique to a non linear objective function.

\begin{remark}
Without losing any generality, we can consider a null interest rate $r$ and a null drift term $\eta$.  
The choice $r = 0$ is not restrictive, indeed the case $r \ne 0$ can be recovered by modifying the drift term $\eta$ in the dynamics
of the risky security and adding $r$ to wealth growth rate \citep[see e.g.][]{Cartea}.
Furthermore we can choose $\eta=0$: the case with a drift $\eta$ not equal to zero can be recovered by a simple 
translation of the process $X_t$.
\end{remark}

The problem is then equivalent to the selection of the optimal trading strategy
for an investor that has the possibility to choose the amount to invest in a risk-free asset identically equal to $1$ and in a risky security whose
log-price is
\begin{equation}
dX_t= - \kappa \, X_t \,dt + \sigma\,dB_t \, . 
\label{eq:OU}
\end{equation}

Let us first consider a trading strategy where the investor can have a long position in the risky asset. 
The stop-loss is activated when the log-price reaches a level $L$ (lower than zero in this case) selling the risky asset.
This is the typical situation that is considered in practice by hedge funds in reversal statistical arbitrage strategies: 
the long position is closed out when the risky security has a given percentage drop to the mean 
level.

Once the stop-loss $L$ is fixed,
the ansatz for selecting the trading strategy is identified by an entry band $D$ and an exit band $U$, 
as in \citet{bertram1}.
In the entry band $D$ the investor buys the security and sells it in the exit band $U$.
The entry band $D$ 
is lower 
than zero, the mean-reverting value for $X_t$ in (\ref{eq:OU}), because the investor bets on the mean reversion of the log-price and
$D$ has to be larger than the stop-loss $L$. 

\bigskip

Once $L$ has been chosen, the  trading strategy is defined by the following steps that generalize the ones in 
\citet{bertram1} to the case with a stop-loss:
\begin{enumerate}
	\item the position is opened when the process reaches $D$, buying the security;
	\item if the process arrives at $U$ without having reached the stop-loss $L$ before, 
  	the position is closed in $U$ with a profit (positive scenario). 
	Then the investor has to wait for the process $X_t$ to return to $D$ to re-open the position;
	\item if the process arrives at $L$ before reaching $U$, the investor is forced to close out the position in $L$ with a loss (negative scenario). Then, also in this case, 
 	he will wait for the process to return to $D$ to re-open the position.
	 Trade length is the time interval between two subsequent trades.
\end{enumerate}

\begin{figure}[h]
	\begin{center}
		\includegraphics[scale=1]{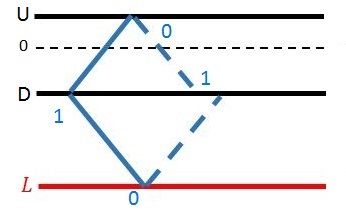}
		\caption{\small Sketch of the strategy with the trading bands and the stop-loss. The investor opens the long position (1) in $D$ and 
		closes it (0) in $U$ or in $L$ (continuous blue lines). He has to wait that the process $X_t$ returns in $D$ to re-open the position (dashed lines). Trade length is the time between two trades.}
		\label{fig:1level}
	\end{center} 
\end{figure}

Figure \ref{fig:1level} shows the relationship between the trading bands and the stop-loss. 
Each trade starts with the log-price equal to $D$ and ends the first time $X_t$ exits the channel $(L,U)$. 

\bigskip  

Since many of the quantities related to the trading strategy (e.g. the return) depend on the time between two subsequent trades in the strategy,
let us carefully define them.

\begin{definition} [First-Exit-Time and First-Passage-Time]
The First-Exit-Time from the channel $(L,U)$ starting in $D$, with $L < D < U$, is a random variables defined as
\be
	\tau_e  \coloneqq  \inf \{t:X_t \notin (L,U) | X_0=D\}
\label{eq: def FET}
\en
while $\tau^+_{e}$ ($\tau^-_{e}$)  represents the FET in the positive (negative) scenario
\be
\left\{
\begin{array} {lcl} 
	\tau^+_{e} & \coloneqq & \inf \{t:X_t \notin (L,U), X_t \ge U | X_0=D\}\\
	\tau^-_{e} & \coloneqq & \inf \{t:X_t \notin (L,U), X_t \le L |X_0=D\} \, .\\
\end{array}
\right.
\label{eq: def FET positive}
\en 
Also First-Passage-Times  play a relevant role in trade length, since one has to include also the time waited to enter again in a new trade. 
They can be defined as
\ben
\left\{
\begin{array} {lcl} 
 \tau_p (U \rightsquigarrow D)   &  \coloneqq & \inf\{t:X_t \le D |X_0=U\} \\
 \tau_p (L \rightsquigarrow D) &  \coloneqq & \inf\{t:X_t \ge D |X_0=L\} \, .
\end{array}
\right.
\enn 
\label{def: times}
\end{definition}


In the two scenarios described above, for a continuous process $X_t$, trade lengths are
\be
\left\{
\begin{array} {lcl} 
 \tau^+ & := & \tau^+_{e} + \tau_p (U \rightsquigarrow D) \\
 \tau^- & := & \tau^-_{e} + \tau_p (L \rightsquigarrow D)\,
\end{array}
\right.
\label{eq: trade length}
\en
where $\tau^+$ ($\tau^-$) represents the trade length in the positive (negative) scenario. 

\bigskip

As already mentioned in the introduction, 
the investment strategy is not realized  only once but it is reproduced iteratively using a self-financing portfolio.
At time $t_i$, corresponding to the $i^{th}$ trade, the investor diversifies his entire available wealth in the two assets
with $f_i$ the fraction of wealth invested in $t_i$ in the risky asset.
In this trading strategy, if price dynamics is continuous,
the wealth between the $i^{th}$ trade and the next one are related via
\be
 W_{t_{i+1}} = W_{t_{i}} \left[ 1 + f_i \, v\right]
\label{eq: capital dynamics}
\en
where $v$ is a random variable that can assume two values
\be
\begin{array} {ccc}
\left\{
\begin{array} {lcl}
 v^+ & \coloneqq & e^{U-D-c}-1 \\
 v^- & \coloneqq & e^{L-D-c}-1  
\end{array}
\right. & {\rm with \, probability} &
\begin{array} {lcl}
p^+ & \coloneqq \mathbb{P} \left[ X_{\tau_e^+}=U| X_0 = D \right] \\
p^- &  \coloneqq \mathbb{P} \left[ X_{\tau_e^-}=L| X_0 = D  \right]
\end{array}
\end{array} \, ,
\label{eq: multip v}
\en
where we have considered the total transaction cost $c$ associated to each trade.
The fraction $f_i$ is positive because 
we are considering an investor who enters in a long position;
when $f_i>1$ the investor is leveraging his investment borrowing some more capital.
The quantity $f_i$ models leverage at time $t_i$.

Because security dynamics parameters and entry/exit bands are time independent,
in the case investor decision does not depend on 
his total wealth, every time the process reaches $D$ the investor repeats the strategy
allocating always the same $f$ to the risky security. 

\bigskip

Summing up, the following conditions should be satisfied, 
 in order to have a well posed financial problem for the trading strategy of interest: 
\be 
i) \, U-D > c 
\ \ \ \ \ \  ii) \, D-L > 0  
\ \ \ \ \ \  iii) \, f \ge 0 \, .
\label{eq: conditions}
\en
The first condition is equivalent to avoid trading inside the transaction costs; 
it  is equivalent to state that $v^+ > 0 $.
The second condition imposes that the stop-loss $L$ is always lower than the entry level $D$: it implies that
 $ v^- < 0 $.
The other condition states 
that we are considering a long position in the risky security.


We can now state two of the main results of this study, that hold for a generic continuous price dynamics.

\begin{proposition}
If price dynamics is continuous,
for a given stop-loss level $L$, the long-run return   as a function of the leverage $f$ and the trading bands $D$ and $U$ is:
	\begin{equation}
	\mu(f, D,U| L)=\frac{p^+ \,  \ln(1+f v^+)+p^-  \ln(1+f v^-)}{p^+\E[\tau^+]+p^-\E[\tau^-]}\, 
	\label{eq:mu_1}
	\end{equation}
\label{prop: mu}
\end{proposition}
{\it Proof. }
See Appendix A $\clubsuit$

\bigskip
Let us observe that the denominator of (\refeq{eq:mu_1}) is the expected trade length.
In particular, 
as shown in the next section, 
when $X_t$ is OU (\ref{eq:OU}) the quantities $p^+$, $\E[\tau^+]$ and $\E[\tau^-]$ are analytical functions of $D, U$ and $L$.

\bigskip

\begin{proposition}
If the first two moments of the trade length are finite,
the variance of the return, $\V [\mu_t]$, is $O(1/t)$ when $t$ goes to infinity. 
\label{prop: mu variance}
\end{proposition}
{\it Proof.} 
See Appendix A $\clubsuit$

\bigskip

The above proposition states that, 
for the described strategy,
the return $\mu_t$  
 is an almost sure quantity in an infinite horizon. Stated differently the Sharpe Ratio of the strategy, 
defined as the expected excess return (w.r.t. the risk-free rate $r$) over 
 return's standard deviation, goes to infinity in the long run.  

\bigskip

Before analysing the optimal return let us introduce the probability for the ``fair  game", that takes into account also the transaction costs paid.

\bigskip

\begin{definition}[Fair probability]
We define the following two quantities:
\bea
q^+  \coloneqq \frac{v^-}{v^- - v^+}\,, \ \ \ \  & {\displaystyle q^-  \coloneqq \frac{v^+}{v^+ - v^-}} = 1 - q^+\,.
\ena 
\end{definition}

\bigskip

If condition i) in {\rm(\refeq{eq: conditions})} holds then $q^+ \in (0,1)$ and $ \mathbb{Q} \coloneqq \{ q^+, q^-\}$  is a probability set.
It is possible to interpret $ \mathbb{Q}$ as the fair probability set associated to the random variable $v$, i.e.
\[
 q^+ v^+ + q^- v^- = 0 \, .
\]

\bigskip

We can divide the optimization problem in two steps: first, we find the optimal leverage $f^*$ with $D$ and $U$ fixed, then we search the optimal trading bands.

\bigskip

\begin{lemma}
If price dynamics is continuous, the optimal leverage $f^*$ satisfies:
\begin{equation}
\label{eq:f_opt}
f^*=\left\{
\begin{array}{ll}
{\displaystyle - \frac{p^+v^+ + p^-v^-}{v^+ v^-} } & {\rm if\;} p^+ > q^+ \\[4mm]
0 & {\rm otherwise}
\end{array}
\right.  \, .
\end{equation}
\end{lemma}
{\it Proof. }
Straightforward after having imposed to the long-run return (\ref{eq:mu_1}) the first order condition ${\displaystyle \frac{\partial \mu}{\partial f}|_{f=f^*}=0}$. 
The optimal leverage $f^*$ satisfies condition iii) in (\ref{eq: conditions}) when $p^+ > q^+ $; otherwise $f^*=0 \,\, \clubsuit$

\bigskip

The above lemma states that it is worth investing a non-null amount in the strategy ($f^* > 0$) only if, for the chosen bands $D$ and $U$, the  
``game'' is unfair and it is mean reverting, i.e. the probability of the positive scenario $p^+$ is larger than the fair probability $q^+$.
We could impose the trading strategy to a generic continuous price dynamics; however it makes sense only 
when dynamics is mean-reverting.

\bigskip

\begin{corollary}
The long-run return {\rm (\refeq{eq:mu_1})} with the optimal leverage $f^*$ is 
\begin{equation}
\mu (f^*,D,U| L) = \left\{
\begin{array}{ll}
{\displaystyle
\dfrac{p^+\ln(\dfrac{p^+}{q^+})+p^-\ln(\dfrac{p^-}{q^-})}{p^+\E[\tau^+]+p^-\E[\tau^-]} } & {\rm if\;} p^+ > q^+ \\[4mm]
0 & {\rm otherwise}
\end{array}
\right.  \, ,
\label{eq:mu_entropy}
\end{equation}
i.e. it is the Kullback-Leibler divergence between probability $\mathbb{P}\coloneqq \{ p^+, p^-\}$ and 
$\mathbb{Q}$, normalized by the expected time between two subsequent trades, when $p^+ > q^+$ and it is zero otherwise.
\end{corollary}
{\it Proof. }
Straightforward from direct substitution of the optimal leverage $f^*$ (\ref{eq:f_opt}) $\; \clubsuit$

\bigskip

We can observe that the numerator of (\refeq{eq:mu_entropy}) is very simple and
we demonstrate in the next section that also the denominator can be expressed in an elementary way for an OU.
The long-run return   (\refeq{eq:mu_entropy}), given the stop-loss $L$,
depends only from the trading bands $U$ and $D$. It can be maximized 
in order to find the optimal values for the trading bands, as we show in a detailed example in section {\bf 4}.

\begin{remark}
When a short position in the risky asset is allowed,  the long-run return   of the strategy 
can be further increased. 
Given the symmetry of OU dynamics (\ref{eq:OU}) the investor can consider also the reverted strategy doubling the long-run return: 
he opens a short position in $-D>0$ and closes it buying back the risky security in $-U$ (profit) or in $-L$ (loss).
\end{remark} 

\section{First-Exit-Time moment of the Ornstein-Uhlenbeck process}

In previous section we have shown that the optimal long-run return   depends
on the probabilities to reach trading bands and on some expected FPTs and FETs,
for any continuous price process.
When $X_t$ follows OU dynamics (\ref{eq:OU}), the expression for expected FPT and probability to reach endpoints can be found in the literature; in this section we show how to compute expected FET in terms of elementary functions.

\begin{remark}
It is possible to rescale time $t$ and log-prices $X_t$ in {\rm (\ref{eq:OU})}, 
measuring time in $1/\kappa$ units, and $X_t$ in $\sigma/\sqrt{\kappa}$ units, obtaining an equivalent dynamics 
(in the new units of measure) with  $\kappa = \sigma= 1$. Stated differently,
 trading bands are multiples of the standard deviation $\Sigma := \sigma/\sqrt{2 \, \kappa}$ for 
the Gaussian stationary distribution for the process $X_t$ {\rm(\refeq{eq:OU})};
times (e.g. trade lengths) are multiples of OU characteristic time scale $\theta := 1/\kappa$.
\end{remark}
 
We indicate with lowercase letters  trading bands in $\Sigma$ units, i.e.
\[
D= d \, \Sigma, \qquad U= u \, \Sigma \, .
\]
We can also indicate with $l$ the stop-loss in $\Sigma$ units 
\[
L= l \, \Sigma
\]
with $l$ a negative real number. 
In mathematical terms this choice corresponds
to stop the strategy when the drawdown
reaches a given quantile of the Gaussian stationary distribution. 
A typical choice for $l$ is $- 1.96$ that corresponds to the $2.5\%$ quantile.

\bigskip

In order to compute the expected FET, we consider FET's Laplace transform in \citet{Borodin} 
and the expression for the parabolic cylinder function in \citet{NRS}
\be
D_{-\lambda} (x) = \frac{\sqrt{\pi} \; 2^{- \lambda/2}}{\Gamma(\frac{1 + \lambda}{2})} \; e^{- x^2/4} \; \sum^{\infty}_{k=0} \lambda^k \, \chi_k \left( - \frac{x}{\sqrt{2}} \right) \, , \qquad \lambda \in \Re^+, x \in \Re
\label{eq:ExpansionParabolicCylinder}
\en
with $\Gamma: \Re^+ \to  \Re^+ $ the gamma function and the functions $\{ \chi_k \}_{k \in \mathbb{N}}$ defined via the following iterative relation  
\be
\left\{
\begin{array}{lcll}
\chi_0 (z) &:= & 1 & z \in \Re \\
\chi_i (z)  &:= & { \displaystyle 2 \, \int^z_0 dy \,  e^{y^2}  \int^y_{-\infty} du \,  e^{-u^2} \chi_{i-1} (u) } & z \in \Re, i \in \mathbb{N}^+ \\
\end{array}
\right.
\label{eq:recursive}
\en

In particular, it can be useful to decompose $\chi_i (z)$ functions in even and odd functions, as shown in the following lemma for the first two.

\bigskip

\begin{lemma}
For the first two functions $\{ \chi_i (z) \}_{i=1,2}$ defined in {\rm (\ref{eq:recursive})}, we have
\be
\left\{
\begin{array}{lcl}
\chi_1(z)  & = & \sqrt{\pi} \, \varphi_1 (z) + \psi_1 (z) \\
\chi_2(z)  & = & \sqrt{\pi} \, \left[ \varphi_2 (z) -  \ln 2 \,\, \varphi_1 (z) \right] + \psi_2 (z) 
\end{array} 
\right. \;  z \in \Re
\label{eq:PairOdd}
\en
where $\{ \psi_i \}_{i=1,2}$ are even functions while $\{ \varphi_i \}_{i=1,2}$ are odd functions, whose definitions can be found in the Glossary of Functions.
\end{lemma}

{\it Proof}. See Appendix {\bf A} $\clubsuit$

\bigskip


The quantity of interest is $\tau_e$ in (\refeq{eq: def FET}), the FET starting from $D$ and exiting from the interval $(L,U)$ with $L< D < U$;
in particular we are interested in the FET $ \tau^+_e $  ($ \tau^-_e $) 
when exiting from $U$ ($L$).

The probability to exit from above  \citep[p.548 eq.3.0.4b]{Borodin} is
\be
p^+ = 
\frac{{\rm Erfid}(d,l)}{{\rm Erfid}(u,l)}  = 
\frac{\varphi_1 \left( \frac{d}{\sqrt{2}} \right) - \varphi_1 \left( \frac{l}{\sqrt{2}} \right)}
        {\varphi_1 \left( \frac{u}{\sqrt{2}} \right) - \varphi_1 \left( \frac{l}{\sqrt{2}} \right)} \; ,
\label{eq:prob p}
\en
where the last equality is obtained rewriting the probability as a function of the -previously mentioned- odd function $ \varphi_1 $. 

The main result of this section is stated in the following proposition, where it is provided an explicit expression for the first moment of $\tau^+_e$
for the OU (\ref{eq:OU}).

\bigskip

\begin{proposition}
The first moment of $\tau^+_e$ in {\rm (\ref{eq: def FET positive})} for the OU dynamics {\rm (\ref{eq:OU})} is
\be
\begin{split}
\frac{\E[ \tau^+_e]}{ \theta }  = &  
 	   \frac{\varphi_2 \left( \frac{u}{\sqrt{2}} \right) - \varphi_2 \left( \frac{l}{\sqrt{2}} \right)  -
		\psi_1  \left( \frac{u}{\sqrt{2}} \right) \, \varphi_1 \left( \frac{l}{\sqrt{2}} \right) +
		\psi_1  \left( \frac{l}{\sqrt{2}} \right) \, \varphi_1 \left( \frac{u}{\sqrt{2}} \right)  }
	{\varphi_1 \left( \frac{u}{\sqrt{2}} \right) - \varphi_1 \left( \frac{l}{\sqrt{2}} \right)} -  \\
& \	    \frac{\varphi_2 \left( \frac{d}{\sqrt{2}} \right) - \varphi_2 \left( \frac{l}{\sqrt{2}} \right)  - 
		\psi_1  \left( \frac{d}{\sqrt{2}} \right) \, \varphi_1 \left( \frac{l}{\sqrt{2}} \right) +
		\psi_1  \left( \frac{l}{\sqrt{2}} \right) \, \varphi_1 \left( \frac{d}{\sqrt{2}} \right)  }
	{\varphi_1 \left( \frac{d}{\sqrt{2}} \right) - \varphi_1 \left( \frac{l}{\sqrt{2}} \right)}
\end{split}
\en
where $\{ \psi_i \}_{i=1,2}$ and $\{ \varphi_i \}_{i=1,2}$ are respectively even and odd functions, whose definitions can be found in the Glossary of Functions
and $\theta$ the characteristic time of the OU. 
\label{prop:FET}
\end{proposition}

{\it Proof}. See Appendix {\bf A} $\clubsuit$

\bigskip

The first moment of the First-Passage-Time $\tau_p (D \rightsquigarrow U)$ (with $D<U$) is obtained in the following corollary.

\bigskip

\begin{corollary}
The first moment of the FPT $\tau_p (D \rightsquigarrow U)$ with $D<U$ is
\be
\frac{\E[ \tau_p (D \rightsquigarrow U)]}{ \theta }  =
\sqrt{\pi}
\left[ \varphi_1 \left( \frac{u}{\sqrt{2}} \right) - \varphi_1 \left( \frac{d}{\sqrt{2}} \right)  \right]+
\left[ \psi_1 \left( \frac{u}{\sqrt{2}} \right) - \psi_1 \left( \frac{d}{\sqrt{2}} \right)  \right]
\en
\end{corollary}

{\it Proof}. This result can be proven as shown in \citet{NRS} or as a limit for $L \to -\infty$ of previous proposition $\, \clubsuit$

\bigskip

Similarly, when exiting in $L$, one has  the probability 
\[
p^- =
\frac{{\rm Erfid}(u,d)}{{\rm Erfid}(u,l)} = 
\frac{\varphi_1 \left( \frac{u}{\sqrt{2}} \right) - \varphi_1 \left( \frac{d}{\sqrt{2}} \right)}
        {\varphi_1 \left( \frac{u}{\sqrt{2}} \right) - \varphi_1 \left( \frac{l}{\sqrt{2}} \right)} \, ,
\]
the expected FET $\tau^-_e$ is
\be
\begin{split}
\frac{ \E[ \tau^-_e]}{\theta} =  & 
 	   \frac{\varphi_2 \left( \frac{u}{\sqrt{2}} \right) - \varphi_2 \left( \frac{l}{\sqrt{2}} \right)  -
		\psi_1  \left( \frac{u}{\sqrt{2}} \right) \, \varphi_1 \left( \frac{l}{\sqrt{2}} \right) +
		\psi_1  \left( \frac{l}{\sqrt{2}} \right) \, \varphi_1 \left( \frac{u}{\sqrt{2}} \right)  }
	{\varphi_1 \left( \frac{u}{\sqrt{2}} \right) - \varphi_1 \left( \frac{l}{\sqrt{2}} \right)} -  \\
& \	    \frac{\varphi_2 \left( \frac{u}{\sqrt{2}} \right) - \varphi_2 \left( \frac{d}{\sqrt{2}} \right)  - 
		\psi_1  \left( \frac{u}{\sqrt{2}} \right) \, \varphi_1 \left( \frac{d}{\sqrt{2}} \right) +
		\psi_1  \left( \frac{d}{\sqrt{2}} \right) \, \varphi_1 \left( \frac{u}{\sqrt{2}} \right)  }
	{\varphi_1 \left( \frac{u}{\sqrt{2}} \right) - \varphi_1 \left( \frac{d}{\sqrt{2}} \right)} \, 
\end{split}
\en
and the expected value of FPT $\tau_p (U \rightsquigarrow D)$ with $D < U$ is
\be
\frac{\E[ \tau_p (U \rightsquigarrow D)]}{ \theta }  =
 \sqrt{\pi} \left[ \varphi_1 \left( - \frac{d}{\sqrt{2}} \right) - \varphi_1 \left( - \frac{ u}{\sqrt{2}} \right)  \right] +
 \left[ \psi_1 \left( - \frac{d}{\sqrt{2}} \right) - \psi_1 \left( - \frac{ u}{\sqrt{2}} \right)  \right] \; .
\label{eq:FirstMomentDown}
\en

\bigskip

\begin{remark}
Using the same technique of proposition \ref{prop:FET}, also the following moments of FET can be computed.
In particulary it can be shown that also the variance of $\tau^+_e$ and $\tau^-_e$ are finite, as required by proposition 2.
\end{remark}

\bigskip

Previous results for FETs allow writing the expected trade length in the denominator of {\rm (\ref{eq:mu_1})}  and {\rm (\ref{eq:mu_entropy})} 
as a simple expression when $X_t$ follows OU dynamics
(\ref{eq:OU}), as shown in the following corollary.

\bigskip

\begin{corollary}
The expected trade length for the OU dynamics {\rm (\ref{eq:OU})} is
\[
\frac{p^+ \mathbb{E}[\tau^+] + p^- \mathbb{E}[\tau^-]}{\theta}  =   
 \pi \, \frac{{\rm Erfid}(d,l) \, {\rm Erfid}(u,d)}{{\rm Erfid}(u,l)} \, .
\]
\end{corollary}

{\it Proof}. Straightforward from direct substitution $\, \clubsuit$

\bigskip

In the OU case we can write explicitly long-run return (\refeq{eq:mu_1}), due to the simple expressions for probabilities and expected trade length.
It becomes
\be
\mu(f, D,U| L)=\frac{1}{\pi \, \theta}\left( 
 \frac{ \ln  \left[ 1 + f \left( e^{U-D-c} - 1\right) \right]}{{\rm Erfid}(u,d)} + 
 \frac{ \ln  \left[ 1 + f \left( e^{L-D-c} - 1\right) \right]}{{\rm Erfid}(d,l)} 
 \right)\, .
\label{eq:mu_final}
\en

\bigskip

In previous section we have discussed for a generic continuous price process
the condition for having non trivial trading strategy, i.e. a non-null amount invested in the risky asset.
Let us now comment on the main consequences when $X_t$ follows OU dynamics (\ref{eq:OU}).
As shown in Lemma 1, for a given trading strategy (and then for a given set of bands $D$ and $U$), it is required that 
\[
p^+ > q^+ \Leftrightarrow c <\overline{C} := \ln \left[ 1+ p^+ \left( e^{U- L} - 1\right) \right] - (D-L) \;  .
\]

\begin{figure}[h!]
	\begin{center}
		\includegraphics[scale=0.7]{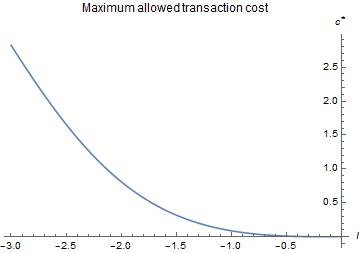}
		\caption{\small  Maximum allowed transaction cost $c^*$ (in $\Sigma$ units) as a function of the stop-loss level $l$ in the range $(-3,0)$.}
\label{fig:maxCost}
	\end{center} 
\end{figure}

In the case of pair trading, the volatility $\Sigma$ takes low values (typically lower than $5 \%$). 
In this case we can consider the first order expansion of 
$\overline{C} =\overline{c} \, \Sigma + O(\Sigma^2)$ with
\[
\overline{c} = p^+ (u-l) - (d-l) \; ,
\] 
where $p^+$, in the OU case, is given by (\ref{eq:prob p}).
In the OU case, it is straightforward to show that $\overline{c}$ is always positive $\forall u \in (d, -l)$. 
Stated differently, every time log-prices follow an OU (with a positive mean reversion $\kappa$) the  ``game" is unfair and mean-reverting ($p^+ > q^+$),
providing that transaction costs $c$ are not too high. 
For any chosen stop-loss level $l$ a quantity of practical interest is $c^*$,
the maximum transaction cost that allows a  non trivial trading strategy,  i.e.
 the maximum of $\overline{c}$ w.r.t. $d$ and $u$.
This quantity $c^*$ is a universal function of the stop-loss level $l$ (i.e. it does not depend on any model parameter) and
in figure \ref{fig:maxCost} we show its plot for $l$ in the range of values relevant for practical purposes.

\bigskip

\begin{remark}
It can be interesting to obtain from {\rm (\refeq{eq:mu_final})} Bertram's long-run return $\mu_B$ as a limit case for a stop-loss $L \to - \infty$.
In this case 
the expected trade length becomes $\pi \, \theta \, {\rm Erfid}(u,d)$ and one gets 
\[
\displaystyle \mu_B (f, D,U) = \frac{\ln \left[ 1 + f \left( e^{U-D-c} - 1\right) \right]}{\pi \, \theta \, {\rm Erfid}(u,d)}\quad ,
\]
\end{remark}
which is an increasing function of $f$, because there are no losses in any scenario. Choosing $f =1$ one gets the well known result
\citep[see e.g.][eq.15]{bertram1} 
\[
\displaystyle \mu_B (f=1, D,U) = \frac{U-D-c}{\pi \, \theta \, {\rm Erfid}(u,d)} \quad .
\]

\section{An application to the energy market}

In this section we show a simple application of the trading strategy with  stop-loss and leverage on empirical data in the energy market.
We consider one of the pairs analyzed in \citet{bucca} that presents a limited seasonality:
the spread between the second future contract of Heating Oil and the sixth future contract of Gas Oil (generally identified as HOc2-LGOc6).

\bigskip

As described in \citet{bucca} the Heating Oil future (HO) is traded in USD per gallon on the New York Mercantile Exchange (NYMEX) with a contract unit of 42,000 gallons, 
while the Gas Oil future (LGO) is traded in USD per metric ton with a contract unit of 100 metric tons
on the Intercontinental Exchange (ICE); we consider a half-an-hour dataset for one year between the 23.04.2015 and the 22.04.2016. Data provider is Thomson-Reuters.
Both markets are round the clock markets, with a 60 minute break between 17:00  and 18:00  New York Time (NYT) of each business day for HO 
(on Sunday market opens at 18:00 NYT) and 
with a two-hour break between 18:00 and 20:00 NYT of each business day for LGO (on Sunday market opens at 20:00 NYT).
We use the classic approach of dividing our dataset into a 9-month In-Sample (IS) period and a 3-month Out-of-Sample (OS) period.
 In the IS period we calibrate OU parameters  
considering only the time interval between 9:00-16:00 NYT when both markets are more liquid;
outside this time slot the volume is significantly smaller. We also determine the optimal trading bands for the strategy.
Strategy's performance is verified OS using round-the-clock prices
(excluding the time interval between 17:00-20:00 NYT, where one of the two markets is closed).
Similarly to \citet{bucca} we consider the ratio ${\cal R}_{t} := p^{(1)}_t/{p^{(2)}_t}$, 
where $p^{(1)}$ represents the mid price of the HO future contract, while $p^{(2)}$ represents the mid price of the LGO future contract.

\begin{table}[h!]
	\caption{Parameters estimated for the HOc2-LGOc6  time-series. We report the estimated values 
and the $95\%$ Confidence Interval (CI) obtained through a bootstrap technique with $10^4$ samples. Conversions are done to ensure time-series are quoted consistently in dollars per barrel. 
}
	\centering
	\begin{tabular}{ccc}
		\toprule
		Parameter	& Estimated Values  & CI \\
		\midrule
		$\hat \kappa$   & 18.51 &  (10.89, 37.55)  \\
		$\hat \eta \, (\%)$   & -0.94 &   (-1.84, -0.01) \\
		$\hat \sigma  \, (\%)$    & 8.93 &  (8.69, 9.16) \\
		\bottomrule   
	\end{tabular}  
	\label{tab:trad HOc2-LGOc6}
\end{table}

\bigskip

Our dataset is composed by 2,908 and 2,752 observations, respectively, in the IS and in the OS periods. 
We consider two filtering techniques in order to clean the time-series from outliers.
 
Following the same technique described in \citet{Benth08}, we search for extreme outliers of ${\cal R}_{t}$.
Given the lower and upper quartiles, $Q_1$ and $Q_3$, respectively, and the interquartile range $IQR := Q_3 - Q_1$, an observation is considered an outlier if it is smaller than 
$Q_1 - 3 \times IQR$, or larger than $Q_3 + 3 \times IQR$. Following this rule, no outlier is detected.

Then we filter out antipersistent outliers that can influence the strategy: 
an observation is classified in this way if the difference between two subsequent ratios is, in absolute value, larger 
than the interquartile range, i.e. $|{\cal R}_{t} - {\cal R}_{t-1}| > IQR$, 
and next observation ${\cal R}_{t+1}$ recovers at least the $95\%$ of $IQR$.
Following this technique we remove $1$ outlier both in IS (on the 3rd of July 2015 at 12:30 NYT) and in OS time series  (on the 15th of Feb. 2016 at 20:00 NYT).

\bigskip

Parameters are estimated on the IS dataset via maximum likelihood; confidence intervals are obtained via a bootstrap technique  with $10^4$ samples (see e.g. Appendix B for details).
Numerical results are reported in table \ref{tab:trad HOc2-LGOc6}.

\begin{figure}[h!]
	\begin{center}
		\includegraphics[scale=0.7]{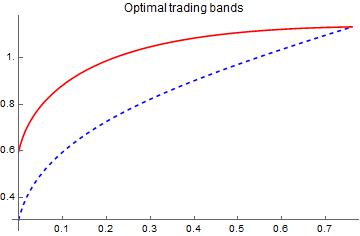}
		\caption{\small Optimal trading bands as a function of transaction costs.
		 We consider a stop-loss $l = - 1.96$, no leverage ($f=1$), and  parameters equal to the estimated ones in Table \ref{tab:trad HOc2-LGOc6},
		 in this case $\Sigma= 1.47 \%$. 
	In \citet{bertram1} trading bands $u$ and $d$ are symmetric; 
           instead in presence of a stop-loss the trading bands $u$ and $d$ are not anymore symmetric. 
	The plot shows $|d|$ (red) and $u$ (blue dashed line) as function of the transaction cost (in $\Sigma$ units) in the range $(0, c^* = 0.76)$, where $c^*$ is the maximum allowed transaction cost for the considered stop-loss level $l$, as computed in previous section.}
\label{fig:bertramL_tradingBands}
	\end{center} 
\end{figure}

In a similar way to \citet{bertram1}, we obtain the optimal trading bands maximizing numerically return (\refeq{eq:mu_final}) for the optimal leverage;
we allow also for short positions.
The optimization is immediate due to the simplicity of the expression.
In figure \ref{fig:bertramL_tradingBands}  we show 
the optimal trading bands $D$ and $U$ as a function of the transaction cost $c$ when the stop-loss $l$ is $-1.96$, 
parameters are the ones in Table \ref{tab:trad HOc2-LGOc6}.

\begin{figure}[h!]
	\begin{center}
		\includegraphics[scale=0.3]{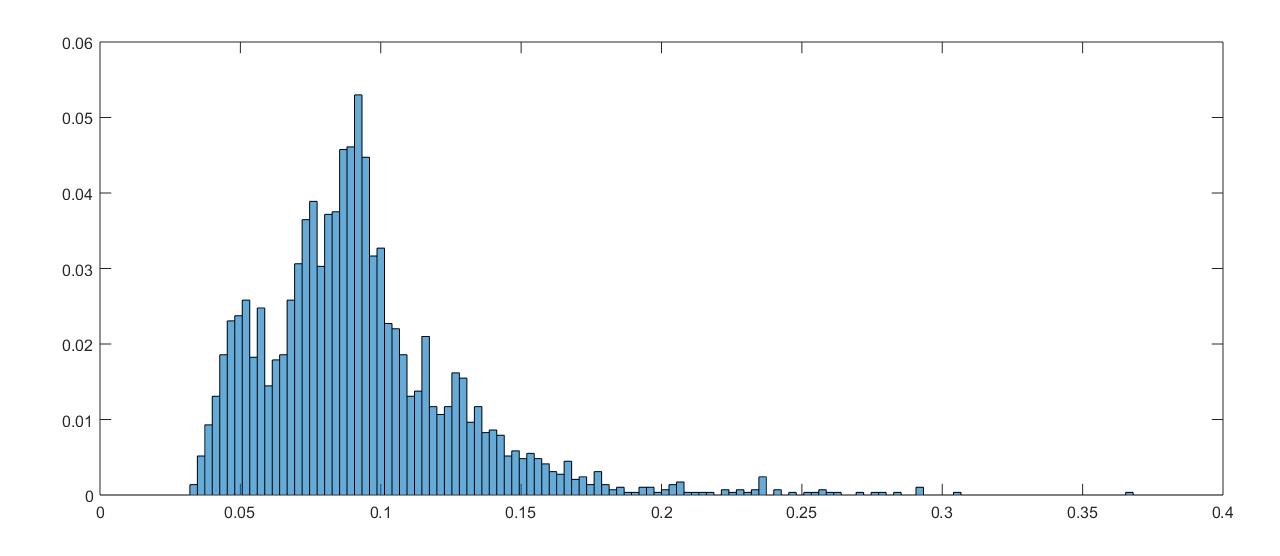}
		\caption{\small Frequency histogram for $c$ in $\Sigma$ units.}
		\label{fig:c_distribution}
	\end{center} 
\end{figure}

Transaction cost parameter $c$ at time $t$ can be estimated from the bid-ask observed in  price time-series as 
\[
c_t =  \sum^2_{i=1} \ln\frac{p^{(i,a)}_t}{p^{(i,b)}_t} 
\]
where $p^{(i,a)}_t$ and $p^{(i,b)}_t$ represent, respectively, the ask and bid price of the $i^{th}$ future with i=1,2.
In figure \ref{fig:c_distribution} we show the empirical distribution of $c$ values; its mean value is $9.33 \% \, \Sigma$.

\bigskip

Optimal leverage is equal to $28.54$, with the $95\%$ Confidence Interval $(15.08,41.97)$,  a  value too large for practical purposes. 
In table \ref{tab:outSample} we show results for optimal bands ($d$ and $u$) and returns with their confidence intervals for different levels of leverage: $1$, $10$ and the optimal leverage. We also report the corresponding OS returns.
We observe that trading strategy results are significant in the long run even introducing a stop loss as shown in this study
and when leverage $f$ is suboptimal.

\begin{table}[h!]
	\caption{Bands ($d$ e $u$) calibrated IS and the corresponding IS return with a stop-loss $l=-1.96$; 
	we show values and their $95\%$ Confidence Interval obtained through a bootstrap technique. 
	We consider three levels of leverage: $1$, $10$ and the optimal one. We also report returns obtained in OS dataset.
	}
	\centering
	\begin{tabular}{cccccccc}
		\toprule
		$f$ & $d$ & CI	& $u$ &  CI & $\mu$ &  CI & $\mu_{OS}$ \\
		\midrule
		$1$   & $-0.870$ & $(-0.915,-0.688)$ & $0.581$ & $(0.564,0.603)$ & $0.145$ & $(0.124,0.151)$ & $0.410$ \\
		$10$ & $-0.863$ & $(-0.924,-0.642)$ & $0.447$ & $(0.386,0.521)$ & $1.175$ & $(1.049,1.367)$ & $4.340$  \\
		opt   &  $-1.108$ & $(-1.112,-0.943)$ & $0.302$ & $(0.296,0.326)$ & $1.945$ & $(0.894,3.707)$ & $5.684$ \\
		\bottomrule   
	\end{tabular}  
	\label{tab:outSample}
\end{table}

\section{Conclusions}
In this paper we develop a statistical arbitrage trading strategy in presence of stop-loss and leverage for a mean reverting security. 

We have first  introduced stop-loss and leverage within the modeling framework presented by \citet{bertram1}.
For any given leverage level $L$, we write analytically the optimal investment strategy and the associated long-run return; 
in order to reach this task we derive, for an OU, the expected First-Exit-Time from a channel in a closed form.
In this case we can deduce a simple analytical expression  for the long-run (\ref{eq:mu_final}).

We have also shown an application to a pair in futures' energy market, considering 12 months half-an-hour time series.   
Optimal bands and leverage can be easily computed also  imposing a stop-loss.
Optimal leverage is too large for practical purposes, but Out-the-Sample results show that the proposed trading strategy
provides significant long-run returns even considering a suboptimal leverage.  

\section*{Acknowledgements}
We thank all participants to the 
Conference on the Mathematics of Energy Markets 
at WPI in Vienna 2016 and to the 
Workshop on Energy Finance in Padova 2016 and in particular
J. Palczewski for useful comments.
The usual disclaimers apply.


\bibliography{Pairs}
\bibliographystyle{tandfx}

\section*{Notation and Glossary of Functions}
\subsection*{Notation}

\begin{table}[h!]
\centering
\begin{tabular}{ll} 
\toprule
\textbf{Symbol} & \textbf{Description}   \\[1ex]
\midrule
	$X_t$ & risky security log-price at time $t$, modeled as an OU \\[1ex] 
	$B_t$ & Brownian motion \\[1ex]
	$c$ & transaction cost parameter, $ c \in \Re^+$ \\[1ex]
	$\mu_t$ & return per unit time up to time $t$ (return) \\[1ex]
	$\mu$ & long-run return \\[1ex]
	$W_t$ & investor's wealth at time $t$ \\[1ex]
	$\kappa, \eta, \sigma$ & OU parameters: mean reversion speed, stationary mean, diffusion parameter \\[1ex]
	$\theta$ &  the characteristic time of the OU dynamics, $\theta := 1/\kappa$, \\[1ex]
	$\Sigma$ & steady-state standard deviation  for the OU  dynamics, $\Sigma := \sigma/\sqrt{2 \, \kappa}$ \\[1ex]
	$\hat{\kappa}, \hat{\eta}, \hat{\sigma}$ & estimated OU parameters via maximum likelihood \\[1ex]
	$f$ & fraction of wealth allocated to the risky security, it models leverage \\[1ex]
	$f^*$ & optimal leverage \\[1ex]
	$U \, (D)$ & upper (lower) trading band \\[1ex]
	$L$ & stop-loss  \\[1ex]
	$l, \, u \, \&  \, d$ & stop-loss, upper \& lower trading bands in $\Sigma$ units \\[1ex]
	$p^+ \, (p^-)$ & transition probability from $D$\ to $U \, (L)$   \\[1ex]
	$q^+ \, (q^-)$ & ``fair" transition probability  from $D$\ to $U \, (L)$  \\[1ex]
	$r$ & risk free rate \\[1ex]
	$\tau$ & trade length \\[1ex]
	$\{ \phi_i \}_{i=1,\ldots,3}$ & first three moments of the trade length $\tau$ \\[1ex]
	$\tau^+ \, (\tau^-)$ & trade length, defined in eq. \eqref{eq: trade length}, in the positive (negative) scenario    \\[1ex]
	$\tau_{e}$ & First-Exit-Time from the channel $(L,U)$ for a process starting in $D$ \\[1ex]
	$\tau_{p} (D \rightsquigarrow U)$ & First-Passage-Time in $U$ for a process starting in $D$ \\[1ex]
	$v^+$ & increase of wealth in the positive scenario, $v^+ := e^{U-D-c}-1$ \\[1ex] 
	$D_\nu(x)$ & cylindric parabolic function, $\nu \in \Re^+, x \in \Re$  \\[1ex]
	$S(\nu,x,y)$ & cylindric parabolic function with two parameters, $\nu \in \Re^+, x,y \in \Re$ \\[1ex]
\bottomrule   
\end{tabular}
\end{table}

\newpage
\subsection*{Abbreviations}

\begin{table}[h!] 
\centering
\begin{tabular}{ll} 
\toprule
\textbf{} & \textbf{Description}   \\[1ex] 
\midrule
	CI & Confidence Interval \\[1ex]
	FET & First-Exit-Time \\[1ex]
	FPT & First-Passage-Time \\[1ex] 
	IS & In-Sample \\[1ex]
	OS & Out-of-Sample \\[1ex]
	OU & Ornstein-Uhlenbeck process \\[1ex] 
	p.d.f. & probability density function \\[1ex]
	w.r.t. & with respect to \\[1ex]
\bottomrule   
\end{tabular}
\end{table}

\subsection*{Glossary of Functions}

The basic functions are \citep{abramowitz}
\[
\begin{array}{lclcl}
\Erf(x) & := & {\displaystyle \frac{2}{\sqrt{\pi}} \int^x_{0} dt \,  e^{-t^2} } & & \\[4mm]
\Erfi(x) & := & {\displaystyle \frac{2}{\sqrt{\pi}} \int^x_{0} dt \,  e^{t^2} } = - i \Erf(i x) & & \\[4mm]
{\rm Erfc}(x)   & := & {\displaystyle \frac{2}{\sqrt{\pi}} \int^{\infty}_x dt \,  e^{-t^2} } = 1- \Erf(x) && \\[4mm]
\Erfid(x,y) & := &  {\displaystyle \frac{\sqrt{2}}{\sqrt{\pi}} \int^x_{y} dt \,  e^{t^2/2} = \Erfi \left( \frac{x}{\sqrt{2}} \right) - \Erfi \left( \frac{y}{\sqrt{2}} \right) }\; . & & \\[4mm]
\end{array}
\]

Similarly to \citet{NRS}, we introduce
the even functions $\{ \psi_i \}_{i=1,2}$ and the  odd functions $\{ \varphi_i \}_{i=1,2}$ that 
decompose the functions $\{ \chi_i \}_{i=1,2}$ defined via the recursive equation (\ref{eq:recursive})

\[
\begin{array}{lclcl}
\varphi_1(x) & := & {\displaystyle  \int^x_{0} dt \,  e^{t^2}  = \frac{\sqrt{\pi}}{2} \Erfi(x) } & = &  {\displaystyle \sum^\infty_{n=0}  \frac{x^{2n+1}}{n! \, (2n+1) } }\\[4mm]
\psi_1(x) & := &  {\displaystyle  2 \int^x_{0} dt \,  e^{t^2} \int^t_{0} du \,  e^{-u^2} = \sqrt{\pi} \int^x_{0} dt \,  e^{t^2} \Erf(t) } & = &  {\displaystyle \sum^\infty_{n=0}  \frac{2^n \, x^{2n+2}}{(2n+1)!! \,(n+1) } }\\[4mm]
\varphi_2(x) & := & {\displaystyle 2  \int^x_{0} dt_2 \,  e^{t_2^2}  \int^{t_2}_{0} du_2 \,  e^{-u_2^2} \int^{u_2}_{0} dt_1 \,  e^{t_1^2} } & = & {\displaystyle \sum^\infty_{n=0}  \frac{x^{2n+3}}{(n+1)! \, (2n+3) } \sum^n_{k=0}  \frac{1}{ 2k+1 }  }\\[4mm]
\psi_2(x) & := &  {\displaystyle  4 \int^x_{0} dt_2 \,  e^{t_2^2} \int^{t_2}_{0} du_2 \,  e^{-u_2^2} \int^{u_2}_{0} dt_1 \,  e^{t_1^2} \int^{t_1}_{0} du_1 \,  e^{-u_1^2} } & = & {\displaystyle
		\sum^\infty_{n=0}  \frac{2^n \, x^{2n+4}}{(2n+3)!! \,(n+2) } \sum^n_{k=0}  \frac{1}{ k+1 } } \; .\\[4mm]
\end{array}
\]

We have also written their Maclaurin series. 

\section*{Appendix A}
In this Appendix we prove  paper's main results.

\bigskip

{\it Proof } {\bf Proposition 1}

Let us define $N_t^+$ ($N_t^-$) as the number of times the process, started in $D$, 
reaches $U$ ($L$) before $L$ ($U$) in the time interval $[0,t]$. 
Let us
consider investor's wealth $W_t$ in the self-financing strategy at time $t$ 
equal to portfolio value immediately after the last trade before $t$ if price dynamics is continuous
\[
{W_t}=W_0 \, (1+f(e^{U-D-c}-1))^{N_t^+}(1+f(e^{L-D-c}-1))^{N_t^-}= W_0  \, e^{\mu_t \, t}\,.
\]

Then, for a particular realization of the process, we have
\begin{equation}
\mu_t=\frac{N_t^+ \alpha^+ + N_t^- \alpha^-}{t}\, ,
\label{eq: mu_t}
\end{equation}
with 
\ben
\left\{
	\begin{array} {lcl}
	\alpha^+ & \coloneqq & \ln(1+f v^+) \\
	\alpha^- & \coloneqq & \ln(1+f v^-) \, .
	\end{array}
\right.
\enn
We can also rewrite the time $t$ as the sum of the trade length, obtaining:
\begin{equation*}
		\mu_t  =\frac{N_t^+\alpha^+ + N_t^-\alpha^-}{\sum_{i=1}^{N_t^+}\tau_{i}^+ + \sum_{i=1}^{N_t^-}\tau_{i}^-} 
		 = \frac{\frac{N_t^+}{N_t^++N_t^-}\alpha^+  +\frac{N_t^-}{N_t^++N_t^-}\alpha^-}{\frac{N_t^+}{N_t^++N_t^-}\frac{\sum_{i=1}^{N_t^+}\tau_{i}^+}{N_t^+} + \frac{N_t^-}{N_t^++N_t^-}\frac{\sum_{i=1}^{N_t^-}\tau_{i}^-}{N_t^-}}\,.
\end{equation*}

The long-run return   $\mu$ is defined as
the limit for $t \to \infty$ of previous expression; for the law of large numbers the proposition is proven 
after observing that
\[
\begin{array}{rcl}
	{\displaystyle \frac{N_t^+}{N_t^++N_t^-} } & \xrightarrow[]{a.s.} & p^+\,,\\
	{\displaystyle \frac{N_t^-}{N_t^++N_t^-}  } & \xrightarrow[]{a.s.} & p^- \,,\\
	{\displaystyle \frac{\sum_{i=1}^{N_t^+}\tau_{i}^+}{N_t^+} } & \xrightarrow[]{a.s.} & \E[\tau^+]\,, \\
	{\displaystyle \frac{\sum_{i=1}^{N_t^-}\tau_{i}^-}{N_t^-}     } & \xrightarrow[]{a.s.} & \E[\tau^-]\qquad  \, \clubsuit
\end{array}
\]

\bigskip

{\it Proof } {\bf Proposition 2}

We compute the expected value and the variance of $\mu_t$, given by equation (\ref{eq: mu_t}).
Let us observe that $N_t^+ = N_t \, \nu_t$, where $N_t$ is an ordinary renewal process and $\nu_t$ the frequency of OU processes that reach $U$ before $L$ in a time interval $t$; 
the interarrival time (trade length in our case) is described by the p.d.f.  \citep[see e.g.][p.353]{Cox}
\[
f(\tau) = p^+ f^+(\tau) + p^- f^-(\tau) 
\]
where $ f^+(\tau) $ ( $ f^-(\tau) $ ) is the p.d.f. of the interarrival time in the positive (negative) scenario.
Given $N_t$, the frequency $\nu_t$ follows a binomial distribution. 

For an ordinary renewal process with finite first three moments of the interarrival time, the following relations hold  \citep[see e.g.][p.345]{Cox}
\[
\left\{
\begin{array}{lcl}
	\E [N_t] & =  & {\displaystyle \frac{1}{\phi_1} \, t + \frac{\phi_2^2 - \phi_1^2}{2 \phi_1^2} + O\left(\frac{1}{t}\right) } \\[2mm]
	\mathbb{V} [N_t] & = & {\displaystyle \frac{\phi_2^2}{\phi_1^3} \, t + \left( \frac{1}{12} + \frac{5 \phi_2^4}{4 \phi_1^4} - \frac{2 \phi_3}{3 \phi_1^3} \right) +  o\left(1\right) } \\
\end{array}
\right. \, ,
\] 
where $\{\phi_i \}_{i=1,\ldots, 3}$ are the first three moments of the interarrival time, i.e. 
$\phi_1 := \E[\tau], \phi^2_2 := \mathbb{V} [ \tau], \phi_3 : = \E[ (\tau - \phi_1)^3] $.    

We can rewrite (\ref{eq: mu_t}) as 
\[
\mu_t = \frac{  (\alpha^+ - \alpha^-) \,  N^+_t +  \alpha^-  \,  N_t   }{t}
\]
and using conditional expectation we get
\[
 \E [\mu_t] = \frac{1}{t} \left\{ (\alpha^+ - \alpha^-) \, \E \left[ \E [ N^+_t | N_t] \right] +  \alpha^-  \, \E [ N_t ] \right\} = \frac{\overline{\alpha}}{t} \, \E [ N_t ] =  
 \mu   + \overline{\alpha} \, \frac{\phi_2^2 - \phi_1^2}{2 \phi_1^2} \, \frac{1}{t} +
 O\left(\frac{1}{t}\right)^2
\] 
where $\overline{\alpha} := p^+ \, \alpha^+ + (1-p^+) \, \alpha^- $ and $\mu = \overline{\alpha}/\phi_1$ is the long-run return.

Using the Law of total variance \citep[see e.g.][]{Weiss} we obtain
\[
\begin{split}
\mathbb{V} [\mu_t] & = 
        \frac{1}{t^2} \left\{ \E \left[  \V [ (\alpha^+ - \alpha^-) \,  N^+_t +  \alpha^-  \,  N_t  | N_t] \right] +   \, \V [ \E [ \mu_t   | N_t] ] \right\} = \\
 & = \frac{1}{t^2} \left\{ \E \left[  (\alpha^+ - \alpha^-)^2 \, \V [   N^+_t  | N_t] \right] +  \V [ \overline{\alpha} \,  N_t ]   \right\} = \\
& = \frac{1}{t^2} \left\{  (\alpha^+ - \alpha^-)^2 \, p^+ \, (1-p^+) \, \E [ N_t]  +  \overline{\alpha}^2 \, \V [ N_t ]   \right\} = \\
 &  = \frac{1}{t} \left\{ (\alpha^+ - \alpha^-)^2 \, p^+ \, (1-p^+) \,\frac{1}{\phi_1} +   \overline{\alpha}^2 \, \frac{\phi^2_2}{\phi^3_1} \right\} + \\
&  \;\;\;\;  \frac{1}{t^2} \left\{ (\alpha^+ - \alpha^-)^2 \, p^+ \, (1-p^+) \, \frac{\phi_2^2 - \phi_1^2}{2 \phi_1^2} +  
	 \overline{\alpha}^2 \, \left( \frac{1}{12} + \frac{5 \phi_2^4}{4 \phi_1^4} - \frac{2 \phi_3}{3 \phi_1^3} \right) \right\} + o\left(\frac{1}{t}\right)^2 \, .
\end{split}
\]
 If only the first two moments of the interarrival time are finite, we obtain only the leading term $O\left( 1/t \right)$ in above expansion, proving the proposition.

For practical purposes it can be useful to relate the first two moments of $\tau$ with the corresponding quantities in the positive and negative scenario. After straightforward computations one gets
\[
\left\{
\begin{array}{lcl}
	\phi_1 = p^+ \, \phi^+_1 + (1-p^+) \, \phi^-_1  \\[2mm]
 	\phi^2_2 = p^+ \,  (1-p^+)  \, (\phi^+_1 - \phi^-_1)^2 + p^+ \, (\phi^+_2)^2 + (1-p^+) \, (\phi^-_2)^2 \qquad \, \clubsuit
\end{array}
\right.
\]

\bigskip

{\it Proof } {\bf Lemma 2}

Straightforward for $ \chi_1(z) $. 
For $\chi_2(z)$ one should 
\[
\begin{split}
\chi_2(z) =  & {\displaystyle  4 \int^x_{0} dt_2 \,  e^{t_2^2} \int^{t_2}_{-\infty} du_2 \,  e^{-u_2^2} \int^{u_2}_{0} dt_1 \,  e^{t_1^2} \int^{t_1}_{-\infty} du_1 \,  e^{-u_1^2} } = \\[4mm]
 & \sqrt{\pi} \, \varphi_2 (z) + \psi_2 (z) +  {\displaystyle  4 \int^x_{0} dt_2 \,  e^{t_2^2}  \int^{0}_{-\infty} du_2 \,  e^{-u_2^2} \int^{u_2}_{0} dt_1 \,  e^{t_1^2} \int^{t_1}_{-\infty} du_1 \,  e^{-u_1^2} } 
\end{split}
\]
which gives the required result after observing that  the last three integrals can be computed explicitly
\[
I := \int^{0}_{-\infty} du_2 \,  e^{-u_2^2} \int^{u_2}_{0} dt_1 \,  e^{t_1^2} \int^{t_1}_{-\infty} du_1 \,  e^{-u_1^2} = - \frac{\sqrt{\pi}}{4} \ln 2 \; .
\]
As a matter of fact, having defined 
\[
\begin{array}{lcl}
g(u_2)  & := & - {\displaystyle \int^{\infty}_{u_2} dt \,  e^{-t^2} =  - \frac{\sqrt{\pi}}{2} \, {\rm Erfc} (u_2)  } \\[4mm]
f(u_2)  & := &  {\displaystyle \int^{u_2}_{0} dt_1 \,  e^{t_1^2} \, {\rm Erfc} (t_1) }
\end{array}
\]
the quantity $I$ can be written as
\[
\begin{split}
 I  & = {\displaystyle -  \int^{\infty}_{0} du_2 \,  e^{-u_2^2} \int^{u_2}_{0} dt_1 \,  e^{t_1^2} \int^{\infty}_{t_1} du_1 \,  e^{-u_1^2}  } = 
	- \frac{\sqrt{\pi}}{2} \int^{\infty}_{0} du_2 \,  e^{-u_2^2} \int^{u_2}_{0} dt_1 \,  e^{t_1^2} \, {\rm Erfc} (t_1) = \\[4mm]
 & = - {\displaystyle \frac{\sqrt{\pi}}{2} \int^{\infty}_{0} du_2 \, g'(u_2) \, f(u_2) } = 
	{\displaystyle \frac{\sqrt{\pi}}{2} \int^{\infty}_{0} du_2 \, g(u_2) \, f'(u_2) } = \\[4mm]
 &  = - \left( \frac{\sqrt{\pi}}{2} \right)^2 \int^{\infty}_{0} du_2  \, {\rm Erfc} (u_2)^2 \, e^{u_2^2}   
\end{split}
\]
where we have integrated by parts and recognized tabled integral (4.7.6) in \citep[][p.14]{NgGeller} $\; \clubsuit$

\bigskip  

{\it Proof } {\bf Proposition 3}

This proposition can be proven starting from the Laplace transform for the FET $\tau^+_e$ \citep[p. 548 eq.3.0.5b]{Borodin} 
\[
 \E[e^{-s \, \tau^+_e}| X_{\tau^+_e}=U] 
	= \frac{\E[e^{-s \, \tau^+_e}; X_{\tau^+_e}=U]}{\mathbb{P}[ X_{\tau^+_e}=U]} 
	=: {\cal L}^+_e[ \lambda | U]  = \frac{1}{p^+} \,
	\frac{S(\lambda,d, l)}
	{S(\lambda,u,l)} \, ,
\]
with $\lambda := s \, \theta$ 
and $S(\lambda,x,y)$ the parabolic cylinder function with two parameters \citep[see e.g.][p.647]{Borodin}
\begin{equation*}
	S(\lambda,x,y)=\frac{\Gamma(\lambda)}{\pi}e^{(x^2+y^2)/4}[D_{-\lambda}(-x)D_{-\lambda}(y)-D_{-\lambda}(x)D_{-\lambda}(-y)] \quad
	\lambda \in \Re^+ \; {\rm and} \; x,y \in \Re\, .
\end{equation*}

The Laplace Transform becomes
\[
 {\cal L}^+_e[ \lambda | U] = \frac{1}{p^+} \, \displaystyle \frac
 	{e^{(d^2 + l^2)/4} [D_{- \lambda}(-d)D_{- \lambda}(l)-D_{- \lambda}(d)D_{- \lambda}(-l)] }
	{e^{(u^2+l^2)/4} [D_{- \lambda}(-u)D_{- \lambda}(l)-D_{- \lambda}(u)D_{- \lambda}(-l)]} \, . 
\]
and using equation (\ref{eq:ExpansionParabolicCylinder}) 
\[
\displaystyle 
{\cal L}^+_e[ \lambda | U] = \frac{1}{p^+} \, \frac
 { \sum^{\infty}_{k=0} \lambda^k \, \chi_k \left( \frac{d}{\sqrt{2}} \right) \,  \sum^{\infty}_{i=0} \lambda^i \, \chi_i \left( - \frac{l}{\sqrt{2}} \right) -
    \sum^{\infty}_{k=0} \lambda^k \, \chi_k \left( - \frac{d}{\sqrt{2}} \right) \,  \sum^{\infty}_{i=0} \lambda^i \, \chi_i \left(  \frac{l}{\sqrt{2}} \right) }
 { \sum^{\infty}_{k=0} \lambda^k \, \chi_k \left( \frac{u}{\sqrt{2}} \right) \,  \sum^{\infty}_{i=0} \lambda^i \, \chi_i \left( - \frac{l}{\sqrt{2}} \right) -
    \sum^{\infty}_{k=0} \lambda^k \, \chi_k \left( - \frac{u}{\sqrt{2}} \right) \,  \sum^{\infty}_{i=0} \lambda^i \, \chi_i \left( \frac{l}{\sqrt{2}} \right) } \, .
\]

The expected value of $\tau^+_e$ is related to the first derivative in zero (with opposite sign) of the Laplace transform
\ben
\E[ \tau^+_e]  =  - \theta \left. \frac{\partial }{\partial \lambda} {\cal L}^+_e [\lambda|U] \right|_{\lambda=0} =
   \theta \left[ \xi (U, L)  - \xi (D, L) \right]
\enn
with
\ben
\begin{split}
\xi (U, L) := &  \frac{1}
	{\left( \chi_1 \left( \frac{u}{\sqrt{2}} \right) - \chi_1 \left( - \frac{u}{\sqrt{2}} \right)  \right)
	       - \left( \chi_1 \left( \frac{l}{\sqrt{2}} \right) - \chi_1 \left( -  \frac{l}{\sqrt{2}} \right)  \right)} \\
& 
  \left\{ \left( \chi_2 \left( \frac{u}{\sqrt{2}} \right) - \chi_2 \left( - \frac{u}{\sqrt{2}} \right)  \right)
	       - \left( \chi_2 \left( \frac{l}{\sqrt{2}} \right) - \chi_2 \left( -  \frac{l}{\sqrt{2}} \right)  \right)  + \right. \\
& \left.	\chi_1  \left( \frac{u}{\sqrt{2}} \right) \, \chi_1 \left( - \frac{l}{\sqrt{2}} \right) -
		\chi_1  \left( - \frac{u}{\sqrt{2}} \right) \, \chi_1 \left( \frac{l}{\sqrt{2}} \right) \right\} \, \, .
\end{split}
\enn

Using the decomposition of $\{ \chi_i \}_{i=1,2}$ in odd and even functions, after a straightforward computation one obtains the expected FET, proving the proposition $\; \clubsuit$

\section*{Appendix B}

We briefly recall the statistical techniques we use for the estimation of parameters and their statistical accuracy.

\subsection*{Maximum likelihood estimation (MLE)}

Let us consider the OU process (\refeq{eq:OU complete}) with $\eta$ not equal to zero as, 
in general, time series have a stationary mean different from zero.

In order to estimate the three parameters $\kappa, \eta, \sigma$ we consider a maximum likelihood estimation
\citep[see e.g.][]{YuPhillips}. 

The conditional density of $X_{t_i} = x_i$ at time $t_i$ given $X_{t_{i-1}}= x_{i-1}$ is:
\begin{equation}
\label{eq:OU_density}
f(x_i|x_{i-1};\kappa,\eta,\sigma)=\frac{1}{\sqrt{2\pi\varsigma^2}}\,\exp\biggl\{-\frac{(x_i-x_{i-1}e^{-\kappa \Delta t}-\eta(1-e^{-\kappa \Delta t}))^2}{2\varsigma^2}  \biggr\} \, ,
\end{equation}
where $\Delta t=t_i-t_{i-1}$ and 
\begin{equation*}
	\varsigma^2=\sigma^2\frac{1-e^{-2\kappa \Delta t}}{2\kappa}\, .
\end{equation*}

We maximize the likelihood or equivalently the log-likelihood: 
\begin{equation*}
	\label{eq:logL}
	\ln {\cal L}(\kappa,\eta,\sigma|\{x_i\}_{i=0,\ldots,{\cal N}})=-\frac{\cal N}{2}\ln(2\pi) - \frac{{\cal N}}{2} \ln \varsigma^2 +
 \sum_{i=1}^{{\cal N}} \frac{(x_i-x_{i-1}e^{-\kappa \Delta t}-\eta(1-e^{-\kappa \Delta t}))^2}{\varsigma^2} \, .
\end{equation*}
Maximum likelihood estimators are given by $(\hat{\kappa},\hat{\eta},\hat{\sigma})=\arg\max_{\kappa,\eta,\sigma} \ln {\cal  L}(\kappa,\eta,\sigma)$.

\subsection*{Bootstrap}

In order to check parameters significance it is possible to use a bootstrap technique \citep[see e.g.][]{efron86}. 
It corresponds to follow the steps:
\begin{enumerate}
	\item estimate via MLE the values of $\{ \hat{\kappa},\hat{\eta}, \hat{\sigma} \}$ of the time-series with ${\cal {\cal N}}$ values;
	\item simulate the OU process with parameters $\{\hat{\kappa},\hat{\eta}, \hat{\sigma}\}$ $M$ times (samples)  with ${\cal N}$ values each sample;
	\item estimate a new set of parameters $\{ \hat{\kappa}_i,\hat{\eta}_i, \hat{\sigma}_i \}$  via MLE on each $i^{th}$ sample with $i=1, \ldots, M$; 
	\item given the empirical distributions for the three parameters determine the confidence intervals for each parameter.
\end{enumerate} 

In the paper we consider $M = 10^4$ samples.
	
\end{document}